\documentclass[conference]{IEEEtran}
\IEEEoverridecommandlockouts
\usepackage{cite}
\usepackage{amsmath,amssymb,amsfonts}
\usepackage{algorithmic}
\usepackage{graphicx}
\usepackage{textcomp}
\def\BibTeX{{\rm B\kern-.05em{\sc i\kern-.025em b}\kern-.08em
    T\kern-.1667em\lower.7ex\hbox{E}\kern-.125emX}}

\usepackage[]{graphicx}
\graphicspath{{../pdf/}{../jpeg/}}
\DeclareGraphicsExtensions{.pdf,.jpeg,.png}

\usepackage{threeparttable}
\usepackage{tablefootnote}
\usepackage{multirow}
\usepackage{amsmath,graphicx}
\usepackage{pifont}
\usepackage{fancyhdr}
\usepackage{wasysym} 
\usepackage{diagbox} 
\usepackage{hyperref} 
\usepackage{cleveref}
\usepackage{booktabs} 
\usepackage[table,xcdraw]{xcolor} 
\usepackage{color, soul} 
\usepackage{booktabs} 

\usepackage[linesnumbered,ruled]{algorithm2e} 

\begin{document}

\title{Grey Rhino Warning: IPv6 is Becoming Fertile Ground for Reflection Amplification Attacks
}

\author{
\IEEEauthorblockN{Ling Hu, Tao Yang, Yu Pang, Bingnan Hou, Zhiping Cai, Bo Yu} \\
\IEEEauthorblockA{\textit{National University of Defense Technology, College of Computer Science and Technology}}
}

\maketitle

\begin{abstract}
Distributed Denial-of-Service (DDoS) attacks represent a cost-effective and potent threat to network stability. While extensively studied in IPv4 networks, DDoS implications in IPv6 remain underexplored. The vast IPv6 address space renders brute-force scanning and amplifier testing for all active addresses impractical. Innovatively, this work investigates AS-level vulnerabilities to reflection amplification attacks in IPv6. 

One prerequisite for amplification presence is that it is located in a vulnerable autonomous system (AS) without inbound source address validation (ISAV) deployment. 
Hence, the analysis focuses on two critical aspects: global detection of ISAV deployment and identification of amplifiers within vulnerable ASes. 
Specifically, we develop a methodology combining \textit{ICMP Time Exceeded} mechanisms for ISAV detection, employ IPv6 address scanning for amplifier identification, and utilize dual vantage points for amplification verification.

Experimental results reveal that 4,460 ASes (61.36\% of measured networks) lack ISAV deployment. Through scanning approximately 47M active addresses, we have identified reflection amplifiers in 3,507 ASes. The analysis demonstrates that current IPv6 networks are fertile grounds for reflection amplification attacks, alarming network security.

\end{abstract}

\begin{IEEEkeywords}
reflection amplification, ISAV, IPv6, security
\end{IEEEkeywords}

\section{Introduction}
\label{sec: introduction}
Reflection amplification attacks remain a potent DDoS technique, where spoofed requests to vulnerable servers (e.g., DNS/NTP) generate up to $557 \times$ amplified traffic \cite{SNMP}. The 2018 Memcached attack on GitHub \cite{Memcached_github} (peaking at 1.35 Tbps) demonstrated this threat. As IPv4 defenses improve, attackers are shifting focus to IPv6 networks \cite{DDoS_ipv6}. With 35.67\% global adoption \cite{statistics_akamai}, IPv6's vast address space provides abundant amplifiers, as seen in recent IPv6 DDoS attacks against Neustar \cite{DDoS_ipv6}. This transition creates new security challenges in our new IPv6 landscape.

\textbf{Key Observations.}
While the ongoing battle between attackers and defenders regarding reflection attacks in IPv4 has persisted for years \cite{DNSBomb, TsuKing, TsuNAME}, there is still a notable gap in comprehensive measurements of reflection attacks across the entire IPv6 network. Based on existing research, we can identify \textbf{three critical elements relevant to reflection amplification attacks} that apply to both IPv4 and IPv6.

1. \textbf{ASes, where attackers reside, lack OSAV}: The lack of outbound source address validation (OSAV) \cite{OSAV} in autonomous systems (ASes) where attackers are located allows these ASes to ignore the legitimacy of source addresses for outbound traffic, in turn facilitating source address spoofing.

2. \textbf{ASes housing amplifiers lack ISAV}: Amplifiers often exist within ASes that do not implement inbound source address validation (ISAV) \cite{ISAV2}. This failure means these ASes do not verify whether source addresses of incoming traffic are legitimate according to network topology, allowing spoofed traffic to traverse their networks.

3. \textbf{Servers acting as amplifiers have vulnerable protocols}: Hosts utilized as amplifiers often provide weak services that are not properly configured, such as the Network Time Protocol (NTP) \cite{NTP}. These corresponding protocols \cite{NTP, SNMP} can generate response traffic that is orders of magnitude greater—often tens or even hundreds of times—than the initial query. Broad experiments \cite{TsuKing, TsuNAME} have confirmed that hosts with these vulnerable protocols can be exploited as amplifiers to launch effective DDoS attacks.

\textbf{Challenges.}
Given that the deployment detection of OSAV has been well-established \cite{CAIDA}, current vulnerability analysis in the IPv6 environment faces significant limitations primarily due to the challenges in detecting ISAV and identifying weak reflection points. This situation presents two main challenges.

\textbf{Challenge 1: Efficient Detection of ISAV Deployment at a Global Scale.} Traditional methods \cite{SAV, CAIDA_spoofer, SAV_deployment} rely on internal network volunteers or DNS resolvers, which are often impractical. IVANTAGE \cite{Your_router} utilizes a local probe through ICMP rate limiting mechanisms \cite{ICMP6}, but \textit{requires sending a significant volume of packets in a short time}, which is inefficient, and can produce false positives by ignoring intermediate ASes. Finding a more efficient and accurate detection method under resource constraints is a key challenge.

\textbf{Challenge 2: Identification of Amplifiers in Nearly Infinite Search Space.} IPv6 has a $2^{96}$ times larger address space than IPv4, making exhaustive scanning impractical \cite{entire_scan}. Furthermore, the conditions for forming amplifiers are even more stringent, as these hosts must be capable of reflecting TCP or UDP traffic, rather than \textit{merely responding to ICMP traffic} \cite{Your_router}. The question of how to efficiently identify reflection points within this vast space remains an open challenge.

\textbf{Our Paper.}
Testing every IPv6 address for amplification is infeasible, so we focus on identifying \textbf{AS-level vulnerabilities to reflection amplification attacks in IPv6}, targeting ASes without ISAV as potential amplifier hosts. We propose an efficient measurement scheme using ICMP probing with source address spoofing to detect ISAV, followed by IPv6 scanning to find live hosts and protocol testing to identify amplifiers. All measurements are conducted from two controlled vantage points, and detailed statistics and results are provided. Our code is available at \href{https://gitee.com/ahaBCD/amplification}{https://gitee.com/ahaBCD/amplification}.

\textbf{Contributions.} This paper makes significant contributions to understanding reflection amplification attacks within IPv6 networks by not only revealing potential scenarios for such attacks but also presenting a comprehensive analysis of their extensive characteristics. 

\begin{itemize}
    \item We analyze ISAV deployment at the AS level across the global IPv6 network. Our findings reveal that 4,460 ASes, 61.36\% of all measured ASes, have yet to deploy ISAV. This comprehensive survey underscores the widespread existence of vulnerabilities within IPv6 networks.
    
    \item We assess the feasibility and success rate of executing reflection amplification attacks within identified ASes, using two controlled vantage points. It is shown that devices in 3,507 ASes are capable of reflecting and amplifying traffic, with amplification factors reaching up to $4,267\times$. This highlights the significant risks posed by inadequate security measures in these networks.
    
    \item Finally, we provide a detailed feature analysis of amplifiers across the IPv6 landscape and offer corresponding defense recommendations to enhance network security. 
    
\end{itemize}

\section{Background}
In this section, we review the relevant concepts of reflection amplification and source address validation, concluding with an introduction to our measurement model.

\subsection{Reflection Amplification}
During reflection amplification, an attacker sends a request with a spoofed source address to a server acting as a reflector, enticing the server to respond to a victim. The scale and impact of the attack are determined by factors such as the number of reflectors, the volume of traffic, the duration of the attack, and so on. The amplification factor (AF) can be calculated as described in \cite{SNMP, DNSBomb}:
\begin{small}
\begin{equation}
\begin{split}
AF&=\ \#\ of\ amplifiers\cdot PAF \cdot  SAF \cdot TCF\ ,\\
PAF&=\ \frac{\#\ of\ response\ packets}{\#\ of\ query\ packets}\ ,\\
SAF&=\ \frac{Response\ packet\ size}{Query\ packet\ size}\ ,\\
TCF&=\ \frac{Time\ used\ for\ query}{Time\ used\ for\ response}\ .\\
\end{split}
\end{equation}
\end{small}

Traditional reflection amplification attacks primarily exploit protocols where the response size exceeds that of the query \cite{NTP, SNMP}. Emerging attackers continually innovate, generating larger attack volumes with equal or fewer resources. 
For example, an attacker can make multiple spoofed \textit{monlist} queries to an NTP server to expand the number of interactive clients in advance, prompting more response packets for one query\cite{SNMP}. Additionally, \cite{middleboxes} successfully demonstrated the use of PSH packets to trigger reflection amplification attacks, thereby reducing the size of query packets. The method proposed in \cite{DNSBomb} allows attackers to send multiple DNS requests continuously to increase the time for query accumulating and minimize response time through \textit{query aggregation} and \textit{rapid response return} mechanisms. Moreover, attackers are progressively expanding their reflection types from traditional vulnerable servers to network middleboxes. Research by \cite{middleboxes} indicates non-compliant traffic can induce middleboxes to return substantial traffic volumes without TCP three-way handshake required.

However, these studies predominantly focus on IPv4. This paper focuses on revealing the substantial potential for amplifiers within IPv6 and explores methods to increase amplification factor similarly.

\subsection{Source Address Validation}
Source address validation (SAV) is a critical technology for preventing source address spoofing by verifying the legality of packet source IP address, including inbound source address validation (ISAV) and outbound source address validation (OSAV) \cite{OSAV, ISAV2}. OSAV prevents forwarding packets with spoofed source addresses from within its network, while ISAV restricts the entry of such packets into its network. OSAV deployment detection has matured, while ISAV remains a challenge \cite{CAIDA_spoofer, OSAV, ISAV2, SAV}. Existing ISAV deployment detection methods exhibit significant limitations: \cite{ISAV_detection2} is based on misconfigurations, \cite{ISAV_detection4} relies on passive traffic analysis, while Spoofer \cite{SAV_deployment, CAIDA_spoofer, SAV} demands probes within the network. Additionally, \cite{ISAV2} requires comprehensive network scanning, which is impossible in IPv6, and \cite{OSAV} necessitates the deployment of open DNS resolvers.

The state-of-the-art IVANTAGE \cite{Your_router} relies on ICMP rate limiting, requiring a significant number of packets and risking misclassifying ASes without ISAV deployment as having it due to the ignorance of strict ISAV of the on-path routers. This paper introduces a more efficient and accurate method for assessing ISAV deployment within IPv6 networks, addressing these limitations.

\subsection{Measurement Model}
Our study employs minimally restrictive vantage points capable only of sending spoofed packets (without monitoring or modifying traffic) to ensure broad applicability in assessing AS-level vulnerabilities to reflection attacks. This approach is validated by CAIDA \cite{CAIDA} showing 25.9\% of IPv6 and 21.4\% of IPv4 ASes permit source address spoofing due to lacking OSAV protections. Additionally, the UDP protocol's inherent vulnerability, responding to unverified source addresses, enables attackers to spoof targets and direct amplified traffic floods, with our experiments confirming negligible packet loss during measurements.

\section{Discovering Vulnerable ASes}

\subsection{Detecting ASes without ISAV}
\label{subsec: Detecting Vulnerable ASes}
Autonomous systems (ASes) that have not deployed inbound source address validation (ISAV) are vulnerable to reflection amplification attacks. The key to identifying such ASes is determining whether ISAV filters spoofed-source packets. In the absence of probes within the target AS, the challenge becomes determining whether spoofed-source packets are received by hosts within an AS where they should not usually arrive. In this work, we leverage the Internet Control Message Protocol (ICMP) \cite{ICMP6} to detect such ``receptions'' and employ a cross-verification approach to confirm whether these ``receptions'' are indeed invalid.

Given that we control two vantage points, we could implement a reflection test by one vantage point, which can spoof the source address as the other's, thus triggering hosts within vulnerable ASes to respond to the target. By observing the received traffic at the target, we can infer whether those hosts have ``received'' packets.






To quickly explore ISAV deployment across multiple ASes, we can incrementally set the TTLs of packets, triggering routers along the path from a vantage point to a targeted host to send \textit{ICMP Time Exceeded} messages. Similarly, the host will send \textit{ICMP Destination Unreachable} or \textit{ICMP Echo Reply} packets according to UDP-based or ICMP-based implementations of \textit{traceroute}. The main distinction here is that the vantage point falsifies the source IP address of the probing packets to that of the target. As a result, if the target receives an ICMP error message, it can be determined that the corresponding router or host ``receives'' the spoofed packets.

\subsection{Validation with Traceroute} 
Simply detecting that routers or targeted hosts are ``receiving'' spoofed packets does not necessarily imply that the corresponding ASes have not deployed ISAV. This is because the traffic from both the vantage point and the target may traverse the same routers on their way to the hosts.
From the perspective of the ASes where these routers are located, packets originating from both the vantage point and the target appear to be from the same source in terms of network topology. Under such circumstances, ISAV cannot effectively differentiate or filter the spoofed packets. Therefore, we must rely on \textit{traceroute} data of both the vantage and target to identify truly invalid ``receptions''.

\begin{figure}[htbp]
  \centering  
  \includegraphics[width=0.99\columnwidth]{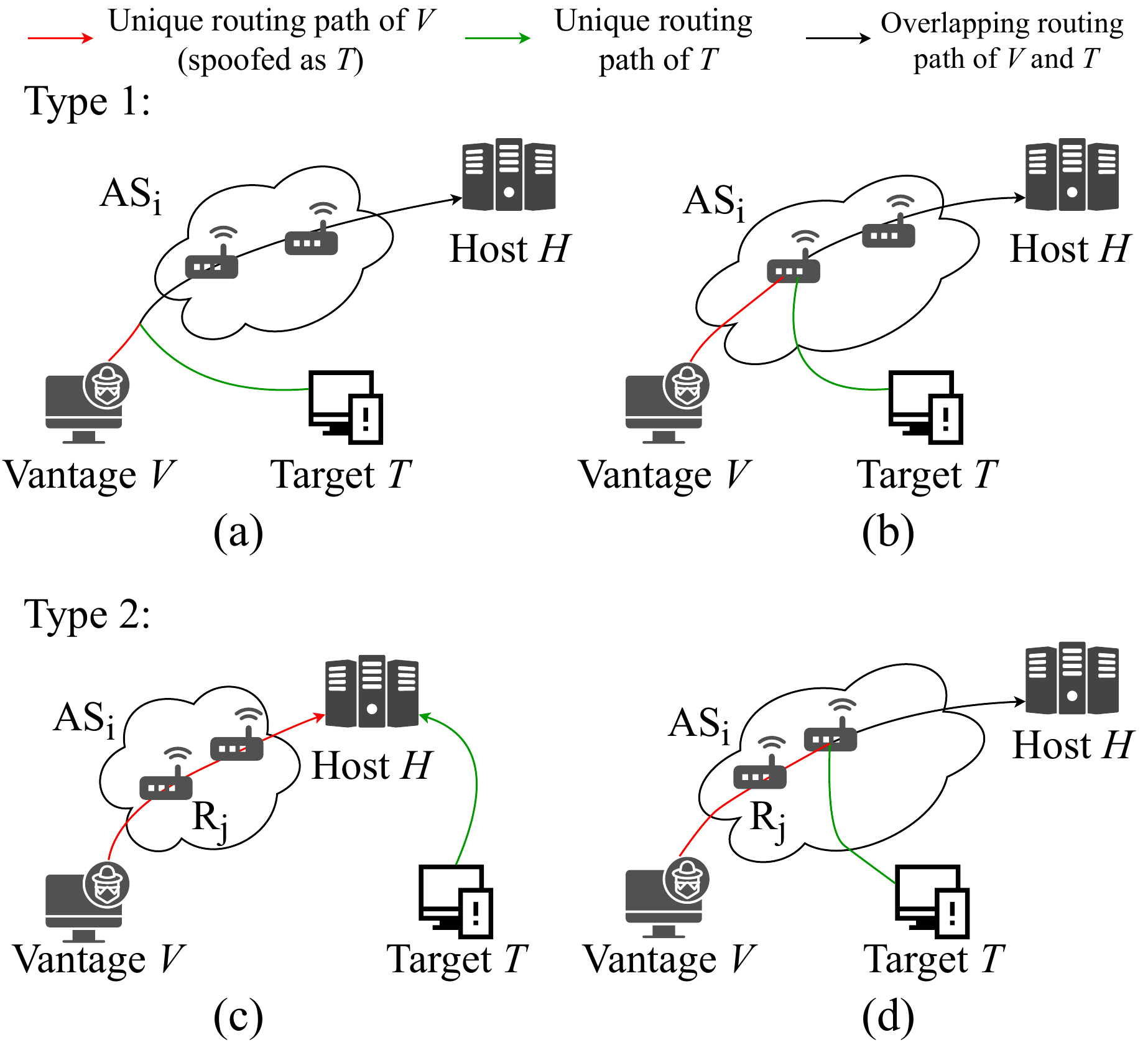}
  \caption{Three types of positional relationships among AS $AS_i$ and routing paths of vantage point $V$ and target $T$ to host $H$. The red, green, and black arrows represent the unique path from $V$ to $H$ with the source address spoofed as $T$, the unique path from $T$ to $H$, and the overlapping path of the two, respectively.}  
  \label{fig: traceroute}       
\end{figure}

As illustrated in Fig.~\ref{fig: traceroute}, this paper concentrates on following two positional relationships among an AS and routing paths of a vantage point and a target to a host.

\begin{itemize}
    \item \textbf{Type 1}: As illustrated in Fig.~\ref{fig: traceroute} (a) and (b), all routers within $AS_i$ are in the path from vantage point $V$ to host $H$ and the path from target $T$ to host $H$. In the case of Fig.~\ref{fig: traceroute} (a), even if $AS_i$ has deployed ISAV, it cannot distinguish spoofed packets that appear to originate from the same source as the target in the network topology. Furthermore, the results of routers in Fig.~\ref{fig: traceroute} (a) and Fig.~\ref{fig: traceroute} (b) ``receiving'' spoofed packets along the route from vantage point $V$ to host $H$ are indistinguishable. Therefore, we cannot definitively determine whether $AS_i$ has deployed ISAV in such scenarios.

    \item \textbf{Type 2}: As shown in Fig.~\ref{fig: traceroute} (c) and (d), there is a router $R_j$ within $AS_i$ that is on the path from vantage point $V$ to host $H$ but not on the path from target $T$ to host $H$. In this scenario, if $R_j$ ``receives'' the spoofed packet, it can be concluded that $AS_i$ has not deployed ISAV.
\end{itemize}

\subsection{Identifying Practical Reflectors}
\label{subsec: Identifying Practical Reflectors}
\textbf{Scanning Method.} 
According to IPASN (October 2024) \cite{IPASN}, there are currently 30,368 ASes enabled in the IPv6 network, covering 225,419 BGP prefixes. To achieve AS-level coverage, we refer to \cite{entire_scan}, performing prefix consolidation and random-byte generation for each remaining BGP prefix to collect IPv6 addresses as targeted hosts (cf. Alg. \ref{alg: Vul_ASes_Discovery} Line 2). Similarly to \cite{Your_router}, we utilize Yarrp \cite{Yarrp} to randomly permute a targeted host × TTL space to trigger reflection at in-path routers in a stateless manner. The corresponding algorithm is presented in Alg. \ref{alg: Vul_ASes_Discovery}.
\begin{algorithm}
\label{alg: Vul_ASes_Discovery}
\caption{Discovering ASes without ISAV}
\small
\KwIn{BGP prefixes $Prefix$}
\KwOut{Vulnerable ASes $Vul\_ASes$}
Initially $Vul\_ASes = [\ ]$\;
$Targeted\_hosts$ = $Random\_generation(Prefixes)$\;
$RT\_vantage$ = Yarrp($Targeted\_hosts$, $vantage$, $vantage$) \;
$RT\_spoof$ = Yarrp($Targeted\_hosts$, $vantage$, $target$) \;
$RT\_target$ = Yarrp($Targeted\_hosts$, $target$, $target$) \;
\For{$host \in Targeted\_hosts$}{
$AS\_host$ = IP\_to\_AS($host$) \;
\If{$AS\_host \notin Vul\_ASes$}{
$Group\_spoof$ = Filter($Routers\_spoof$, $host$) \;
$Group\_target$ = Filter($Router\_target$, $host$) \;
    \For{$router \in Group\_spoof$}{
        \If{$router \notin Group\_target$}{
            $Vul\_ASes$.append($AS\_host$) \;
            break \;
        }
    }
}
}
\end{algorithm}

The vantage point first performs a standard \textit{traceroute} to identify ICMP-responsive routers (Line 3), then spoofs the target's address to detect routers ``receiving'' forged packets (Line 4). By comparing the spoofed and target's \textit{traceroute} results (Lines 5-18), Type 2 vulnerable ASes are identified.
Ideally, routers in ASes before the first ISAV-enabled AS will respond to spoofed packets, enabling multi-AS ISAV detection in one attempt. The vantage point's normal and spoofed paths help identify ISAV-deploying ASes (those responding normally but blocking spoofed traffic), though only the first such AS per host can be confirmed.

Notably, while routers respond with ICMP Time Exceeded messages, this doesn't guarantee UDP-based amplification capability, necessitating further validation for reflection amplification risks.

\textbf{Identifying Vulnerable Servers.} As discussed in Key Observation in Sec~\ref{sec: introduction}, the third crucial condition for hosts to serve as amplifiers is the presence of vulnerable protocols on those hosts. This paper focuses on reflection amplification attacks facilitated by three common UDP services: DNS, NTP, and SNMP \cite{SNMP}. To determine whether these services are available on active addresses collected from IPv6 Hitlists \cite{hou2023search} and proactive target generation scans \cite{yang20226forest}, we use Xmap \cite{periphery_scan} to probe for service presence of DNS, NTP, and SNMP.

\textbf{Reflection Amplification Tests.} We have control over both a vantage point and a target, allowing the vantage point to perform source address spoofing. The vantage point can send service request packets with source addresses spoofed to that of the target, directed at service-providing hosts. Meanwhile, the target listens for the reflected traffic from the hosts. By analyzing this traffic, we can identify various amplifiers and match ASes containing them. This method not only reveals the presence of amplifiers but also enhances our understanding of the vulnerabilities inherent in certain ASes. By systematically probing these services and analyzing the responses, we can effectively map the landscape of potential threats and inform mitigation strategies for vulnerable networks.

\section{Measurement Results}
\subsection{Data Collection}
In October 2024, we conducted five rounds of comprehensive AS-level ISAV deployment detection, using two controlled vantage points in China and Canada. For ASes found lacking ISAV, we collected active addresses followed by service probing and reflection amplification verification. To minimize network congestion, we scanned vulnerable ASes at a rate of 10 Kps for detecting ``receives'' and 1 Kps for cross-validation, with the initial TTL set to 4 to avoid triggering frequent ICMP reports from routers near the vantage point. Meanwhile, service probing was conducted at 10 Kps. To reduce interference with the target and nearby routers, we conducted reflection amplification validation at a rate of only 0.1 Kps. After each scan, we summarized the target response status and the corresponding AS distribution.

\subsection{Are There Actually Amplifiers in ASes without ISAV?}
\textbf{The Deployment of ISAV.}
After conducting five rounds of AS-level scanning, we measured 7,269 ASes. Among them, routers in 5,914 ASes are found to receive spoofed packets. By correlating these results with the target's \textit{traceroute} data, we confirm that 4,460 ASes (61.36\%) are classified as not deploying ISAV. Meanwhile, 1,560 ASes (21.46\%) are found to have deployed ISAV, and the status of the remaining 1,249 ASes (17.18\%) remains uncertain.

Compared to earlier studies, \cite{CAIDA_spoofer} and \cite{SAV} found that 68.6\% of the tested ASes either lacked or only partially deployed ISAV in 2019. Additionally, \cite{Smap} reported that 69.8\% of tested ASes were without ISAV in 2021. Similarly, in 2022, \cite{Your_router} demonstrated that ISAV deployment was either inadequate or only partially completed in 67.37\% of tested ASes, with full deployment occurring in 20.87\%. Our findings are consistent with the results of these studies.

\textbf{Amplifiers.}
We collected approximately 47 million active addresses within these ASes lacking ISAV. Tab.~\ref{tab: amplifiers} presents the service probing and reflection amplification validation results for these addresses. Among the 4,460 ASes lacking ISAV, we successfully identified exploitable reflectors in 3,507 ASes (78.63\%), providing strong evidence to support our hypothesis that ``ASes lacking ISAV are likely to harbor amplifiers.''

\begin{figure}[htbp]
  \centering  
  \includegraphics[width=\columnwidth]{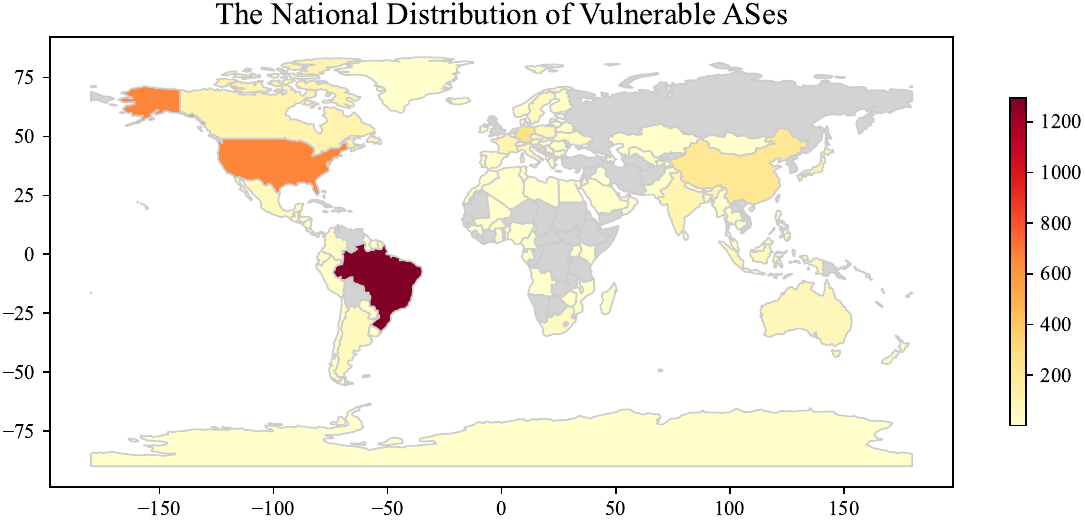}
  \caption{Heatmap of the distribution of vulnerable ASes around the world.}  
  \label{fig: as_country}       
\end{figure}

\textbf{Vulnerable ASes Distribution.}
As shown in Fig.~\ref{fig: as_country}, with data from \cite{maxmind}, we find that the 4,460 ASes without ISAV are distributed across 168 countries. Among them, detection rates for vulnerable ASes in countries and regions like the Middle East and Central Africa are relatively low. This may be because most of these countries and areas have meager IPv6 adoption rates—less than 1\% \cite{statistics_akamai}. 
\begin{table*}[htbp]
\centering
\caption{The number of open servers, verified amplifiers, successful exploitations, and ASes covered for each protocol.}
\label{tab: amplifiers}
\renewcommand\arraystretch{1}
\begin{tabular}{ccccc}
\toprule
Protocols & \# of Hosts Providing Services & \# of Amplifiers & Successful  Exploitations (\%) & \# of ASes Covered\\
\midrule
\rowcolor[HTML]{DFEDDF}
DNS       & 85,952                          & 81,833            & 95.2                          & 596                    \\
NTP       & 34,022                          & 21,691            & 63.8                          & 1,555                   \\
\rowcolor[HTML]{DFEDDF}
SNMPv1   & 3,274                           & 2,803             & 85.6                          & 539                    \\
SNMPv2   & 3,499                           & 3,016             & 86.2                          & 552                    \\
\rowcolor[HTML]{DFEDDF}
SNMPv3   & 53,935                          & 49,952            & 92.6                          & 3,029                   \\
Total     & 158,815                         & 141,006           & 88.8                          & 3,507  
               \\
\bottomrule
\end{tabular}
\end{table*}

It can be concluded that the scope of reflection amplification attacks in IPv6 is relatively wide and warrants caution.

\subsection{How Harmful Amplifiers are in IPv6?}
The key metric for measuring reflection amplification attacks is the amplification factor.
In calculating the amplification factor, we assume a single amplifier, disregard the time used for query and response timing, and include Ethernet headers when determining packet size. The bandwidth amplification factor (BAF) is thus calculated as follows:
\begin{small}
\begin{equation}
    \begin{split}
        BAF&=PAF\cdot SAF \\
           &=\frac{\#\ of\ response\ packets}{\#\ of\ query\ packets}\cdot \frac{Response\ packet\ size}{Query\ packet\ size}.
    \end{split}
\end{equation}
\end{small}

We only select some effective amplifiers to present the amplification performance to minimize the impact on real-world networks.
The measure results and test scenarios for DNS, NTP, and SNMP are summarized in Tab.~\ref{tab: BAF}.

\begin{table*}[htbp]
\centering
\caption{Bandwidth amplification factors for each protocol.}
\label{tab: BAF}
\renewcommand\arraystretch{1}
\begin{tabular}{cccccccc@{}p{0.5\linewidth}@{}|@{}p{0.5\linewidth}@{}}  
\toprule
\multirow{2}{*}{Protocols} & \multirow{2}{*}{\begin{tabular}[c]{@{}c@{}}Query Packet \\ Size (B)\end{tabular}} & \multicolumn{2}{c}{Response Packet Size (B)} & \multicolumn{2}{c}{\# of   Response Packets} & \multicolumn{2}{c}{BAF}   \\
\cmidrule(lr){3-4}
\cmidrule(lr){5-6}
\cmidrule(lr){7-8}
 &      & \multicolumn{1}{c}{Avg}    & Max  & \multicolumn{1}{c}{Avg}   & Max & \multicolumn{1}{c}{Avg}  & Max  \\
\midrule
\rowcolor[HTML]{DFEDDF}
DNS                        & 100                                                                               & 1083.69              & 1429                & 3                     & 3                   & 32.51      & 42.87                              \\

NTP                        & 70                                                                                & 501.15               & 502                 & 92.9                  & 595                 & 665.38     & 4267                                                                                                 \\
\rowcolor[HTML]{DFEDDF}
SNMPv1                    & 96                                                                                & 1297.81              & 1396                & 2.02                  & 3                   & 27.31      & 43.63     
\\
SNMPv2                    & 97                                                                                & 1318.45              & 1505.9              & 8.48                  & 31                  & 115.26     & 481.27    
\\
\rowcolor[HTML]{DFEDDF}
SNMPv3                    & 142                                                                               & 211.03               & 190                 & 3                     & 15                  & 4.46       & 20.07              \\
\bottomrule
\end{tabular}
\end{table*}

\textbf{DNS:} In IPv4, a standard method for conducting reflection amplification using DNS is through name lookups (e.g., A or MX records). With the DNS extension (EDNS0) allowing UDP response packets larger than 512 bytes, the amplification factor generally ranges from 28 to 54 \cite{SNMP}. In our experiment, we test four types of queries: 15, 16, 28, and 255, corresponding to ``MX'', ``TXT'', ``AAAA'', and ``ANY'' query types, respectively. When DNSSEC and EDNS0 are enabled with ``udp payload size'' set to 65535 to break the packet size limit, we achieve an average amplification factor of 32.51 and a maximum of 42.87 for querying Microsoft's ``ANY'' records.

\textbf{NTP:} 
Similar to IPv4, vantages in IPv6 can exploit the \textit{monlist} request supported by NTP servers, which returns up to 100 packets with data on recent clients. 
In our experiment, we do not attempt to maximize the amplification factor for each NTP server to minimize disruption to actual network devices. Instead, we measure the network's existing amplification factors of NTP servers and test an average BAF of 665.38.
In most cases, a \textit{monlist} request can increase request traffic by up to $717.14\times$ while responding with no more than 100 packets. This aligns with the explanation regarding \textit{monlist}.
Surprisingly, some servers can amplify by $4267\times$ with 595 packets responded, which is the highest BAF measured. 

\textbf{SNMP:} 
SNMP has three versions, with SNMPv2 and SNMPv3 supporting the \textit{GetBulk} command for large responses. For SNMPv1, a crafted query packet can achieve an average amplification factor of 27.03 and a maximum of 43.18. SNMPv2, with higher \textit{max\_repetitions}, reaches an average of 115.26 and a peak of 481.27. SNMPv3, however, includes authentication, limiting its amplification factor to a maximum of 20.07 and an average of 4.46. 
As SNMPv3 introduces an authentication mechanism, \textit{GetBulk} is valid for vantages, and the maximum amplification factor for SNMPv3 is limited to 20.07, with an average of only 4.46. Additionally, SNMPv3 exposes router vendor information in error messages \cite{IMC23}.

According to Tab.~\ref{tab: amplifiers}, DNS has the highest exploitation success rate but is concentrated in specific areas. SNMPv3 is widely deployed but has low amplification, while SNMPv2 has higher amplification but less exposure. NTP has a lower exploitation success rate but offers the highest amplification and wider distribution, making it the most ideal amplifier.

\section{Discussion}
\textbf{IP Deactivation.}
IPv6 addresses, especially on mobile devices, frequently change due to network reconnections, leading to short lifespans. Our one-week liveness test found only 4.71\% of previously active addresses remained reachable, confirming rapid turnover. However, this volatility does not affect AS-level vulnerability detection, as reassigned addresses stay within the same AS.
For reflection attacks, IPv6's dynamic nature complicates attacker reconnaissance (requiring repeated scanning) but also enhances anonymity, hindering victim forensics.

\textbf{Countermeasures.} Reflection attacks rely on source IP spoofing. Outbound (OSAV) and inbound (ISAV) source address validation can block spoofed packets: OSAV prevents local-source spoofing, while ISAV filters illegitimate inbound traffic. Despite their importance, many ASes lack full deployment, especially for IPv6. Promoting ISP adoption is crucial, though ISAV only blocks cross-network attacks—internal monitoring remains essential.
What's more, hosts can reduce exposure by restricting access (e.g., DNS servers for internal users only), using authentication (e.g., SNMP passwords), or rate-limiting responses (e.g., DNS RRL). Passive defenses like ACL filtering and DPI can block malicious traffic by IP or content patterns, disrupting attack pathways.

\textbf{Limitations.}
We acknowledge certain limitations in this study. When detecting ASes lacking ISAV, we infer whether routers accept spoofed packets based on ICMP timeout messages received by the target. However, not all routers reliably return ICMP timeout messages. Many gateways are configured by default not to return ICMP messages for security reasons. Consequently, monitoring from the target’s side alone may miss routers receiving spoofed packets.

Additionally, this paper relies on only one probe capable of source address spoofing and one passive listening to obtain all detection results, resulting in limited coverage. Once forged packets are blocked by an AS with ISAV deployed, assessing ISAV deployment in subsequent ASes becomes challenging. 

\textbf{Ethical Consideration.}
We maintain \textbf{anonymity} by only disclosing country-level vulnerable ASes, following established practices. Our scanning uses \textbf{low-impact methods}: standard ICMP Echo requests, distributed traceroute (starting at TTL=4 with randomized targets), and minimal service checks to avoid network disruption. We're coordinating with local network administrators to \textbf{responsibly disclose} identified IPv6 reflection vulnerabilities given their broad implications.

\section{Conclusion}
This paper investigates reflection amplification attacks in IPv6 networks. Using only two probes, we have efficiently assessed the vulnerabilities of ASes and successfully identified reflection amplifiers within these ASes. Our findings confirm the widespread presence of reflection amplifiers in the current IPv6 networks. Additionally, we have analyzed the abuse potential of three UDP-based network protocols in reflection amplification attacks, highlighting this ``grey rhino'' problem.

\section*{Acknowledgment}
This work is supported by the National Natural Science Foundation of China No.62472434 and China Postdoctoral Science Foundation No.2023TQ0089.

\bibliographystyle{IEEEtran}
\bibliography{ref.bib}

\end{document}